\documentclass[onecolumn,showpacs,preprintnumbers,superscriptaddress,amsmath,amssymb]{revtex4}
\usepackage[dvips]{graphicx}
\usepackage{latexsym,amssymb,amsmath,epsfig,bm,times,psfrag}
\usepackage{color}
\usepackage[bookmarksnumbered,bookmarksopen,colorlinks,citecolor=red,linkcolor=blue]{hyperref}
\usepackage{epstopdf}
\newcommand{\Fig}[1]{Fig.~\ref{#1}}

\newcommand{\ie}{\emph{i.e.}}

\newcommand{\eg}{\emph{e.g.}}

\renewcommand{\part}{{\rm part}}
\newcommand{\be}{\begin{equation}}
\newcommand{\ee}{\end{equation}}
\newcommand{\bear}{\begin{eqnarray}}
\newcommand{\eear}{\end{eqnarray}}
\newcommand{\ba}{\begin{array}}
\newcommand{\ea}{\end{array}}

\begin{document}

%\preprint{00-000}

\title{Transverse momentum dependent decorrelation in Pb-Pb collisions at LHC}

\author{De-Xian Wei}
\email{dexianwei@gxust.edu.cn}
\affiliation{School of Science, Guangxi University of Science and Technology, Liuzhou, 545006, China}

\date{\today}% It is always \today, today, but any date may be explicitly specified

\begin{abstract}
Based on A Multi-Phase Transport (AMPT) model simulations, the transverse momentum dependent decorrelation has been studied
in Pb-Pb collisions at $\sqrt{s_{NN}}$= 2.76 and 5.02 TeV, respectively.
It has been found that the mix-order factorization ratio $r_{m, n}$ value deviates significantly from unity in noncentral collisions. Such effect becomes stronger with an increase in the $p_{T}$ difference $p_{T}^{a}-p_{T}^{b}$. These decorrelations are not only between the same order harmonic but also between the different order harmonic, which as a result of the initial fluctuations appear between the different phase spaces.
It has also been found that the correlations involving higher powers of the flow vector yield stronger decorrelation, $r_{m|n;3}<r_{m|n;2}<r_{m|n;1}~(m=2,3; n=2,3,4,5)$, except for the weighted factorization ratio $r_{3|4;k}$.
The breaking phenomenon of these factorization ratios indicated that it provides a possible observation for studying the initial fluctuation properties of heavy-ion collisions.
\end{abstract}

%Use show keys class option if keyword display desired

% PACS, the Physics and Astronomy Classification Scheme.

\maketitle

%=======================================Document Begin=========================================

%======================================= Introduction==========================================
\section{Introduction}
\label{sec:sec1}
The primary goal of ultra-relativistic heavy-ion collisions is to understand the matter properties of Quark-Gluon Plasma (QGP), which is produced in extreme conditions has been predicted by the Quantum Chromodynamics~\cite{Shurya:1980qca}. Anisotropic harmonics flow plays a major role in probing the properties of the QGP at the Relativistic Heavy Ion Collider (RHIC) at BNL~\cite{Abelev:2009lrr} and Large Hadron Collider (LHC) at CERN~\cite{Aad:2012mot}. The realization of higher-order harmonics flow  and its fluctuations~\cite{Gardim:2019ppn}, the correlation between the magnitude and phase of different order harmonics~\cite{Adare:2011moh,Aad:2014moe,Aad:2015mot} and the transverse momentum and pseudorapidity dependence of event plane angles~\cite{Khachatryan:2015eft} have led to a good understanding of the initial fluctuating states and the properties of the strong QGP. Furthermore, the higher-order harmonics ($n~>$ 3) can arise from initial fluctuating anisotropies in the same order harmonic (denoted linear response) or can be driven by lower-order harmonics (denoted nonlinear response)~\cite{Yan:2015nhr,Qian:2016mce,Teaney:2012nli,Acharya:2019nfm}. \par

The experiment indicates that the flow vector fluctuations were observed by the decomposition of Fourier harmonics of the two-particles azimuthal
correlations~\cite{ALICE:2012hdo}. To test the flow vector fluctuations, a useful observable is the factorization ratio, $r_{n,n}$, which encodes the correlations of flow harmonics at different transverse momenta or pseudorapidities~\cite{Khachatryan:2015eft,Gardim:2013bof,Heinz:2013ffa,Aaboud:2017mol,Zhao:2017cfi,Bozek:2018ldm,Bourja:2018fot,Bozek:2018aam,Aad:2020lfd,Chatrchyan:2014soa}. These correlations revealed that the factorization ratio is sensitive to fluctuations in the initial states and not strongly dependent on the viscosity of the system~\cite{Shen:2015saf}. \par

Furthermore, non-unity Person coefficient between different order harmonics has been studied in hydrodynamic~\cite{Niemi:2012edo,Liu:2019pca}.
The different order event plane could be decorrelated as a consequence of initial event-by-event fluctuations~\cite{Xiao:2016foe}. Following these ideas, the main purpose of this paper is to illustrate a particular picture on the initial event-by-event fluctuations driven mixed harmonics decorrelation (denoted mix-order factorization ratio breaking) in Pb-Pb collisions at LHC. \par

%=================================== theory model =========================================
\section{Materials and Methods}
\label{sec:sec2}

One observable to probe the $p_{T}$-dependent flow vector fluctuations is the factorization ratio, $r_{n, n}$~\cite{Gardim:2013bof,Heinz:2013ffa,Bozek:2018aam}. It can be calculated using the two-particle Fourier harmonic by the same order. To test the mixed harmonics flows decorrelation, a mix-order factorization ratio $r_{m, n}$ is expressed as
\begin{eqnarray}\label{fact:st}
r_{m, n}(p_{T}^{a}, p_{T}^{b}) &=& \frac{V_{m,n}(p_{T}^{a},p_{T}^{b})}{\sqrt{V_{m,m}(p_{T}^{a},p_{T}^{a})V_{n,n}(p_{T}^{b},p_{T}^{b})}}, \nonumber \\
V_{m, n}(p_{T}^{a}, p_{T}^{b}) &=& \langle Q_{m}(p_{T}^{a})Q_{n}^{*}(p_{T}^{b})\rangle  \nonumber \\
&=& \langle v_{m}(p_{T}^{a})v_{n}(p_{T}^{b})cos\{m\Psi_{m}(p_{T}^{a})-n\Psi_{n}(p_{T}^{b})\}\rangle.
\end{eqnarray}
Where
\begin{eqnarray}
Q_{m}(p_{T})= |Q_{m}(p_{T})|e^{im\Psi_{m}(p_{T})}=v_{m}(p_{T})e^{im\Psi_{m}(p_{T})}.  \nonumber
\end{eqnarray}
and $V_{m,n}$ is the $m^{th}$- and $n^{th}$-order Fourier harmonic of the two-particle azimuthal correlations of the triggered and associated particles from $p_{T}^{a}$ and $p_{T}^{b}$. Here, $Q_{m}$ and $Q_{n}$ are the $m$-ordr and $n$-order vectors from two different parts of a single event with particles range in a positive or a negative pseudorapidity region, respectively. The $Q_{m}$($Q_{n}$) is the vector for charged particles in the present $p_{T}$ range, and angle brackets denote the average over all events within a given centrality range. To avoid self-correlation, the triggered particles (denoted  $p_{T}^{a}$) are always selected from the positive pseudorapidity region and the associated particles (denoted $p_{T}^{b}$) are from the negative pseudorapidity region.
A psedorapidity gap $|\Delta\eta|>2$ is applied between $p_{T}^{a}$ and $p_{T}^{b}$ to suppress nonflow effects. Correlations between $Q_{m}$($Q_{n}$) of different harmonics represent higher-order correlations which can provide crucial information on the initial-state and its fluctuations of the medium.
For the case $r_{m, n}(p_{T}^{a}, p_{T}^{b}) \leq 1$, it means that the factorization at the transverse momentum $p_{T}^{a}$ and $p_{T}^{b}$ is partially decorrelated. These decorrelations can be due to the flow vector fluctuations both of the flow magnitude and flow angle decorrelate~\cite{Khachatryan:2015eft} generated by initial event-by-event geometry fluctuations.
Flow angle decorrelation means that event-by-event differences in the effective flow angles $\Psi_{m}(p_{T}^{a})$ and $\Psi_{n}(p_{T}^{b})$ at the two different transverse momentum in phase space.
The factor $cos\{m\Psi_{m}(p_{T}^{a})-n\Psi_{n}(p_{T}^{b})\}$ in the numerator of the $r_{m, n}(p_{T}^{a}, p_{T}^{b})$ contributes from the $p_{T}$-dependent fluctuating flow angle that could lead to the ratio breaking~\cite{Xiao:2016foe}. For $v_{m}(p_{T}^{a})$ and $v_{n}(p_{T}^{b})$ in different phase spaces, $a$ and $b$, it will also contribute a flow magnitude fluctuation in event-by-event simulations to the factorization ratio in Eq.~(\ref{fact:st}).
Even though it has been mentioned that flow magnitude and flow angle decorrelate in event-to-event fluctuation by hydrodynamic calculations~\cite{Bozek:2018aam}, this work does not distinguish the decorrelation caused by these two sources, and only considers the overall flow decorrelation.
When $m=n$, it returns the mix-order factorization ratio to be $r_{n, n}(p_{T}^{a}, p_{T}^{b})$.  \par

Correlators of higher powers of the same order flow in two different $p_{T}$ bins have been calculated in hydrodynamic~\cite{Bozek:2018aam}.
Naturally, a mix-order factorization ratio weighted with different powers for $Q_{m}(Q_{n})$ can be defined as
\begin{eqnarray}\label{fact:we}
r_{m|n;k}(p_{T}^{a}, p_{T}^{b}) &=& \frac{V_{m|n;k}(p_{T}^{a},p_{T}^{b})}{\sqrt{V_{m|m;k}(p_{T}^{a},p_{T}^{a})V_{n|n;k}(p_{T}^{b},p_{T}^{b})}}, \nonumber \\
V_{m|n;k}(p_{T}^{a}, p_{T}^{b}) &=& \langle Q_{m}(p_{T}^{a})^{k}Q_{n}^{*}(p_{T}^{b})^{k}\rangle  \nonumber \\
&=& \langle v_{m}^{k}(p_{T}^{a})v_{n}^{k}(p_{T}^{b})cos\{k[m\Psi_{m}(p_{T}^{a})-n\Psi_{n}(p_{T}^{b})]\}\rangle.
\end{eqnarray}
For $k = 1$ one recovers the factorization ratio Eq.~(\ref{fact:st}), $r_{m|n;1}(p_{T}^{a}, p_{T}^{b})=r_{m, n}(p_{T}^{a}, p_{T}^{b})$. \\

In this work, the $p_{T}$-dependent factorization ratios are investigated in Pb-Pb collisions at $\sqrt{s_{NN}}$ = 2.76 and 5.02 TeV for the produced charged particles by the A Multi-Phase Transport (AMPT) model~\cite{Lin:2004amt}, respectively. It takes the specific shear viscosity $\eta/s=0.273$, one is calculated by the Lund string fragmentation parameters in AMPT~\cite{De:2018hri}, \ie, $a=0.5$, $b=0.9$ GeV$^{-2}$,
$\alpha_{s}$ =0.33 and $\mu$ = 3.2 fm$^{-1}$. More thorough details of the AMPT model can be found in Ref.~\cite{Lin:2004amt}. Base on AMPT event-by-event simulation, the factorization ratios are calculated by the scalar-product method~\cite{CMS:2017mom}.  \par

%=======================================calculations and analysis=========================================
\section{Results}
\label{sec:sec3}

\begin{figure*}[tp]
\begin{center}
\includegraphics[width=1.00\textwidth]{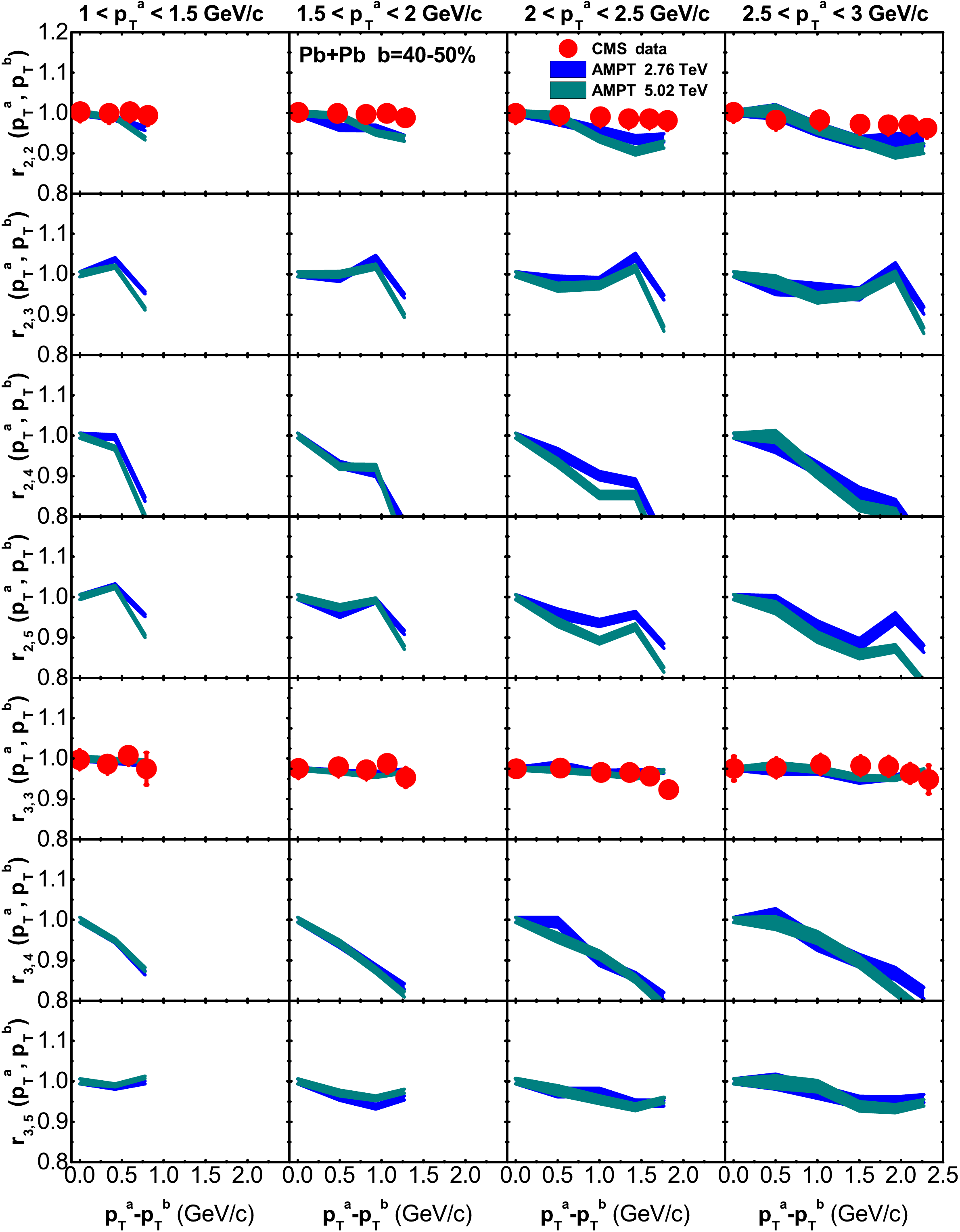}
\caption{(Color online)
Factorization ratio $r_{m, n}$ as a function of the $p_{T}$ difference $p_{T}^{a}-p_{T}^{b}$ in 40-50~\% Pb-Pb collisions at 2.76 and 5.02 TeV from AMPT simulations (colored band), respectively. The simulate results of AMPT are compared with the 2.76 TeV on CMS~\cite{Khachatryan:2015eft} data (red points).
}
\label{fig1}
\end{center}
\end{figure*}

\begin{figure*}[tp]
\begin{center}
\includegraphics[width=0.840\textwidth]{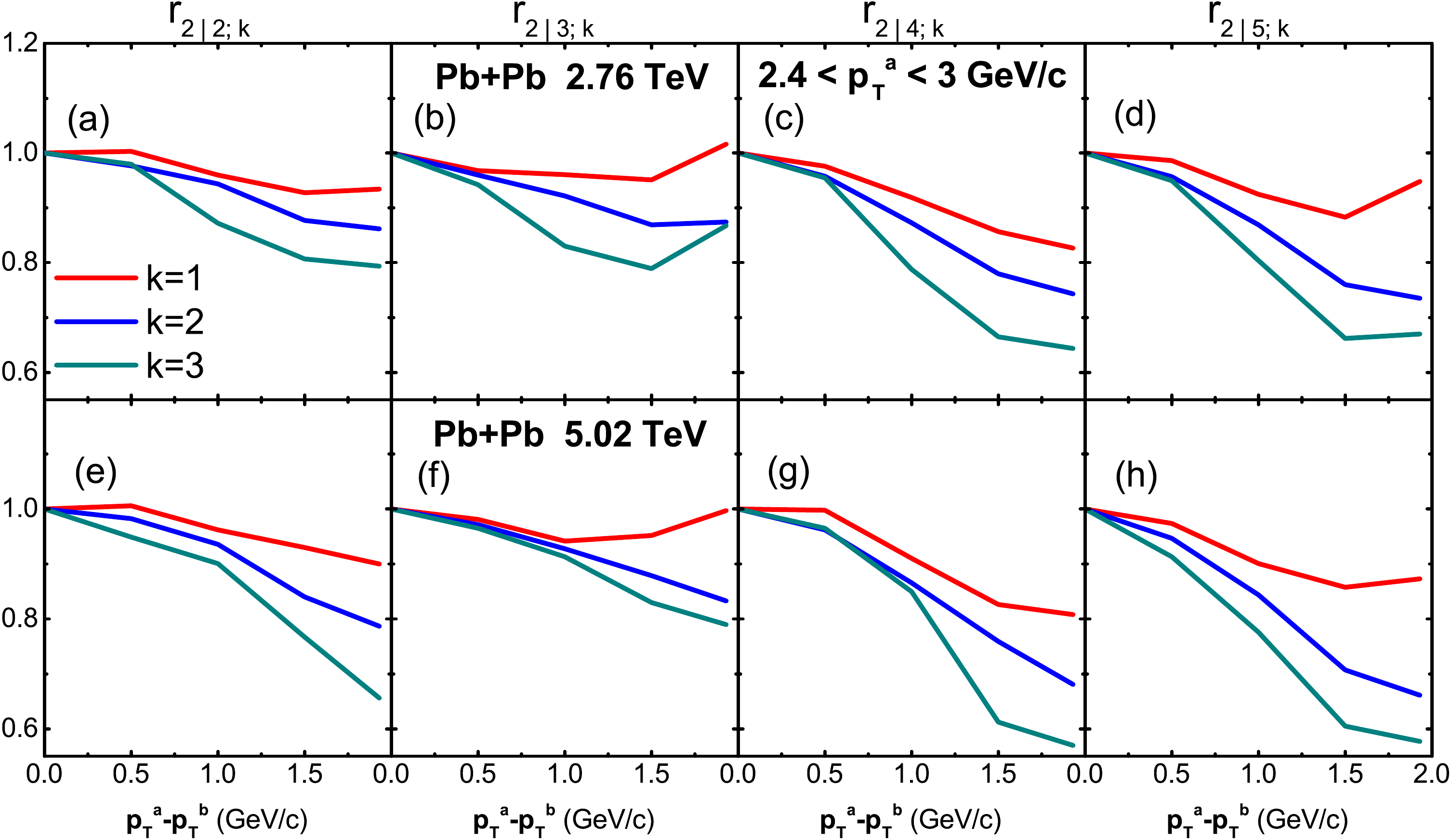}
\caption{(Color online)
The weighted factorization ratio $r_{2|n;k}$ as a function of the $p_{T}$ difference $p_{T}^{a}-p_{T}^{b}$ in 40-50~\% Pb-Pb collisions at 2.76 and 5.02 TeV, respectively. Up panels: for 2.76 TeV Pb-Pb collisions. Down panels: for 5.02 TeV Pb-Pb collisions.
}
\label{fig2a}
\end{center}
\end{figure*}

\begin{figure*}
\begin{center}
\includegraphics[width=0.840\textwidth]{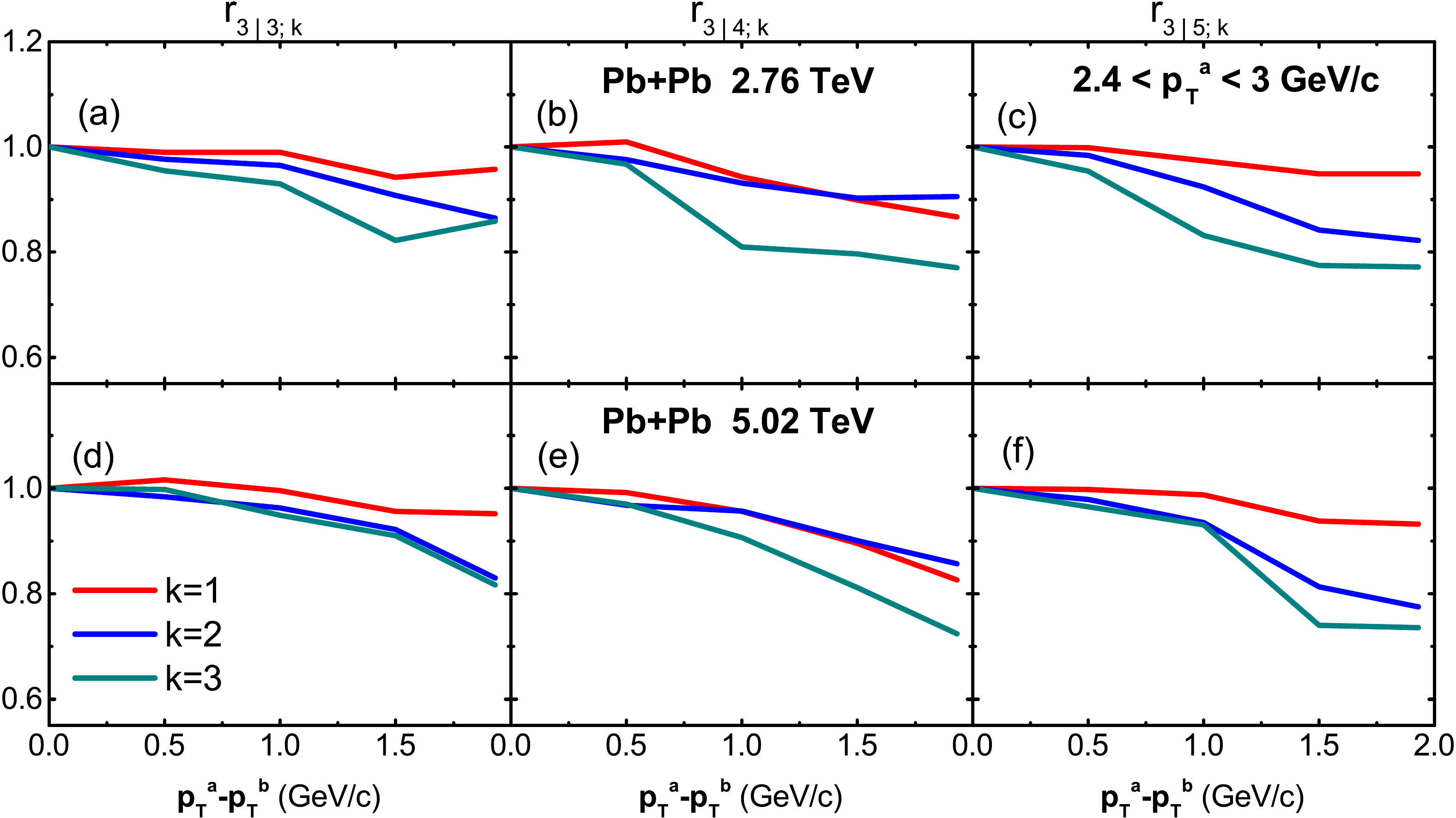}
\caption{(Color online)
Similar distributions as shown in \Fig{fig2a}, but for the weighted factorization ratio $r_{3|n;k}$.
}
\label{fig2b}
\end{center}
\end{figure*}

One study for $p_{T}$-dependent flow vector fluctuations can be via the observable of the factorization ratio, $r_{m, n}$.
Although the specific shear viscosity $\eta/s=0.08$ has completed the calculation of the same order of $r_{m, n}$ ($m=n$) in AMPT~\cite{Acharya:2017sft}, the calculation of different orders of $r_{m, n}$ ($m\neq n$) has not been completed. In addition, hydrodynamic has pointed out that the sensitivity of $r_{m, n}$ to $\eta/s$ is relatively weak~\cite{Shen:2015saf,Acharya:2017sft}. Therefore, this paper focuses on the fluctuation properties rather than the transport properties of QGP medium. In this work, the parameter $\eta/s=0.273$ will be taken.
The results of $r_{m, n}$ are presented in \Fig{fig1} as a function of the $p_{T}$ difference $p_{T}^{a}-p_{T}^{b}$ in 40-50~\% Pb-Pb collisions at 2.76 and 5.02 TeV from AMPT event-by-event simulations (colored band), respectively. AMPT results are compared with the CMS~\cite{Khachatryan:2015eft} data (red points) at 2.76 TeV.
In order to minimize the nonflow effects, a psedorapidity gap $|\Delta\eta|>2$ is taken between the two correlated particles, $a$ and $b$.  It is believed that $|\Delta\eta|>2$ is enough to eliminate the nonflow effects in 40-50~\% collision~\cite{Bourja:2018fot}. If a factorization breaks in the correlation function of the two particles in the case of only the flow effects, it is implied that decorrelation could be caused by the fluctuation of the flow. These flow fluctuations are mainly caused by the initial event-by-event geometry fluctuations in heavy-ion collisions, which is also predicted by hydrodynamic~\cite{Gardim:2013bof,Heinz:2013ffa}.
In \Fig{fig1}, the factorization ratio $r_{m, n}$ significantly deviates from unity in noncentral collisions. This effect becomes stronger with an increase in the $p_{T}$ difference $p_{T}^{a}-p_{T}^{b}$. It is indicated that $p_{T}$-dependent flow vector significantly fluctuates in the presented $p_{T}$ range. From \Fig{fig1}, it shows that the broken effect of the factorization ratio is not only for the same order harmonic but also for the different order harmonic, \eg, $r_{2, 4}$, $r_{2, 5}$ and $r_{3, 5}$. It should be noted that the calculation of $r_{m, n}$ in \Fig{fig1} is based on Eq.~(\ref{fact:st}). There are two contributions in the numerator of Eq.~(\ref{fact:st}), one of which is the contribution of $\{m\Psi_{m}(p_{T}^{a})-n\Psi_{n}(p_{T}^{b})\}$. Event plane $\Psi_{m}(p_{T}^{a})$ and $\Psi_{n}(p_{T}^{b})$ are fluctuating in different phase spaces in AMPT that could produce a factor less than unity. The results of event plane decorrelate based on AMPT simulation have also been mentioned in Ref.~\cite{Xiao:2016foe}. Another part is that $v_{m}(p_{T}^{a})$ and $v_{n}(p_{T}^{b})$ in different phase spaces, $a$ and $b$, will also contribute to a flow magnitude fluctuation. For the event-by-event AMPT simulations, these fluctuations of the two sources will lead to the factorization ratio $r_{m, n}$ breaking, whether $m$ is equal to $n$ or not, the results are shown in \Fig{fig1}. It is also worth noted that there may be additional pseudorapidity-dependent fluctuations of flow angle $\Psi_{m}$ and magnitude $v_{m}$ which are not described in this work. Furthermore, the decorrelations are weakly dependence on the collision energy. \\

Figures \ref{fig2a} and \ref{fig2b} shows the weighted factorization ratio $r_{m|n;k}$ as a function of the $p_{T}$ difference $p_{T}^{a}-p_{T}^{b}$ in 40-50~\% Pb-Pb collisions at 2.76 and 5.02 TeV from AMPT event-by-event simulations, respectively. In \Fig{fig2a}, the up panels are results of $r_{2|n;k}$ for 2.76 TeV on Pb-Pb collisions and the down panels are results of $r_{2|n;k}$ for 5.02 TeV on Pb-Pb collisions, respectively. The charged trigger particles are chosen region in $2.4< p_{T}^{a}< 3.0$~GeV/c. It is found that both ratios not agree with unity over the presented $p_{T}^{a}-p_{T}^{b}$ range. The correlations involving higher powers of the flow vector yield stronger decorrelation, $r_{2|n;3}<r_{2|n;2}<r_{2|n;1}$ as shown in \Fig{fig2a}.
Similar distributions for the weight factorization ratios $r_{3|n;k}$ are also shown in \Fig{fig2b}. From \Fig{fig2b}, the correlations involving higher powers of the flow vector yield also stronger decorrelation, $r_{3|n;3}<r_{3|n;2}<r_{3|n;1}$, except for the weighted factorization ratio $r_{3|4;k}$. As given by Eq.~(\ref{fact:we}), the weighted factorization ratio is also determined by the fluctuations of two different phase spaces, $p_{T}^{a}$ and $p_{T}^{b}$. The fluctuation of the phase space will be amplified after considering the power weight. Therefore, after averaging the events, the weighted factorization ratios will be distributed according to the power weight of fluctuations, \ie, $r_{2|n;3}<r_{2|n;2}<r_{2|n;1}$ and $r_{3|n;3}<r_{3|n;2}<r_{3|n;1}$, except the $r_{3|4;k}$. The $r_{3|4;k}$ do not satisfy the distribution of power weights like other order harmonic factorizations, and the reason is not particularly clear. In addition, hydrodynamic~\cite{Bozek:2018aam} has pointed out that the factorization ratios for higher powers of flow vectors do not factorize into the powers of the basic factorization ratios, \ie, $r_{m|n;k}(p_{T}^{a}, p_{T}^{b})\neq r_{m, n}(p_{T}^{a}, p_{T}^{b})^{k}$. Therefore, it may need to be more careful about reexamination of the fluctuation driven $r_{3|4;k}$.\\

%=======================================summary=========================================
%\section{Summary and discussions}
\section{Summary}
\label{sec:sum}

Base on AMPT event-by-event calculations, this paper has carried out the $p_{T}$-dependent factorization ratio $r_{m, n}$ and weighted factorization ratio $r_{m|n;k}$ in noncentral Pb-Pb collisions at $\sqrt{s_{NN}}$= 2.76 and 5.02 TeV, respectively. The results of AMPT calculations on Pb-Pb systems are compatible with the CMS data within error bars.
By AMPT simulations, it is found that the factorization ratio $r_{m, n}$ value deviates significantly from unity in noncentral collisions. Such effect becomes stronger with an increase in the $p_{T}$ difference $p_{T}^{a}-p_{T}^{b}$. These factorization ratios broken effects are not only for the same order harmonic, but also for the different order harmonic, it indicates that decorrelations appear in the different phase spaces could be caused by event-by-event initial fluctuations.
It has also been found that the correlations involving higher powers of the factorization ratios yield stronger decorrelation, $r_{m|n;3}<r_{m|n;2}<r_{m|n;1}~(m=2,3)$, except for the $r_{3|4;k}$. The breaking phenomenon of these factorization ratios indicates that it provides a possible observation for studying the initial fluctuation properties of heavy-ion collisions.

%%%%%%%%%%%%%%%%%%%%%%%%
\section*{Acknowledgements}
I thank X. G. Huang for very helpful discussion. This work was supported by the Youth Program of Natural Science Foundation of Guangxi (China), with Grant No.~2019GXNSFBA245080, the Special fund for talentes of Guangxi (China), with Grant No.~AD19245157, and also by the Doctor Startup Foundation of Guangxi University of Science and Technology, with Grant No.~19Z19.

%\bibliography{factor_refs}

\begin{thebibliography}{99}
\bibitem{Shurya:1980qca}
      E.~V.~Shurya,
      Phys.\ Rept.\ {\bf 61}, 71 (1980).

\bibitem{Abelev:2009lrr}
      B.~I.~Abelev {\it et al.} [STAR Collaboration],
      Phys.\ Rev.\ C {\bf 80}, 064912 (2009)
      [arXiv:0909.0191 [nucl-ex]].

\bibitem{Aad:2012mot}
      G.~Aad {\it et al.} [ATLAS Collaboration],
      Phys.\ Rev.\ C {\bf 86}, 014907 (2012)
      [arXiv:1203.3087[nucl-ex]].

\bibitem{Gardim:2019ppn}
      F.~Gardim, F.~Grassi, P.~Ishida, M.~Luzum and J.-Y.~Ollitrault,
      Phys.\ Rev.\ C {\bf 100}, 054905 (2019)
      [arXiv:1906.03045 [nucl-th]].

\bibitem{Adare:2011moh}
      A.~Adare {\it et al.} [PHENIX Collaboration],
      Phys.\ Rev.\ Lett.\ {\bf 107}, 252301 (2011)
      [arXiv:1105.3928 [nucl-ex]].

\bibitem{Aad:2014moe}
      G.~Aad {\it et al.} [ATLAS Collaboration],
      Phys.\ Rev.\ C {\bf 90}, 024905 (2014)
      [arXiv:1403.0489 [nucl-ex]].

\bibitem{Aad:2015mot}
      G.~Aad {\it et al.} [ATLAS Collaboration],
      Phys.\ Rev.\ C {\bf 92}, 034903 (2015)
      [arXiv:1504.01289 [nucl-ex]].

\bibitem{Khachatryan:2015eft}
      V.~Khachatryan {\it et al.} [CMS Collaboration],
      Phys.\ Rev.\ C {\bf 92}, 034911 (2015)
      [arXiv:1503.01692 [nucl-ex]].

\bibitem{Yan:2015nhr}
      L.~Yan and J.-Y.~Ollitrault,
      Phys.\ Lett.\ B {\bf 744}, 82 (2015)
      [arXiv:1502.02502 [nucl-th]].

\bibitem{Qian:2016mce}
      J.~Qian, U.~Heinz, and J.~Liu,
      Phys.\ Rev.\ C {\bf 93}, 064901 (2016)
      [arXiv:1602.02813 [nucl-th]].

\bibitem{Teaney:2012nli}
      D.~Teaney and L.~Yan,
      Phys.\ Rev.\ C {\bf 86}, 044908 (2012)
      [arXiv:1206.1905 [nucl-th]].

\bibitem{Acharya:2019nfm}
      S.~Acharya {\it et al.} [ALICE Collaboration],
      JHEP {\bf 06}, 0147 (2020)
      [arXiv:1912.00740 [nucl-ex]].

%\bibitem{Acharya:2020lan}
%      S.~Acharya {\it et al.} [ALICE Collaboration],
%      [arXiv:2002.00633 [nucl-ex]].

\bibitem{ALICE:2012hdo}
      K.~Aamodt {\it et al.} [ALICE Collaboration],
      Phys.\ Lett.\ B {\bf 708}, 249 (2012)
      [arXiv:1109.2501 [nucl-ex]].

\bibitem{Gardim:2013bof}
      F.~Gardim, F.~Grassi, Frederique, M.~Luzum and J.-Y.~Ollitrault,
      Phys.\ Rev.\ C {\bf 87}, 031901 (2013)
      [arXiv:1211.0989 [nucl-th]].

\bibitem{Heinz:2013ffa}
      U.~Heinz, Z.~Qiu and C.~Shen,
      Phys.\ Rev.\ C {\bf 87}, 034913 (2013)
      [arXiv:1302.3535 [nucl-th]].

\bibitem{Aaboud:2017mol}
      M.~Aaboud {\it et al.} [ATLAS Collaboration],
      Eur.\ Phys.\ J.\ C {\bf 78}, 142 (2018)
      [arXiv:1709.02301 [nucl-ex]].

\bibitem{Zhao:2017cfi}
      W.~Zhao, H.~ Xu and H.~ Song,
      Eur.\ Phys.\ J.\ C {\bf 77}, 645 (2017)
      [arXiv:1703.10792 [nucl-th]].

\bibitem{Bozek:2018ldm}
      P.~Bozek and W.~Broniowski,
      Phys.\ Rev.\ C {\bf 97}, 034913 (2018)
      [arXiv:1711.03325 [nucl-th]].

\bibitem{Bourja:2018fot}
      C.~Bourjau,
      [arXiv:1807.05004 [nucl-th]].

\bibitem{Bozek:2018aam}
      P.~Bozek,
      Phys.\ Rev.\ C {\bf 98}, 064906 (2018)
      [arXiv:1808.04248 [nucl-th]].

\bibitem{Chatrchyan:2014soa}
      S.~Chatrchyan {\it et al.} [CMS Collaboration],
      JHEP {\bf 02}, 088 (2014)
      [arXiv:1312.1845 [nucl-ex]].

\bibitem{Aad:2020lfd}
      G.~Aad {\it et al.} [ATLAS Collaboration],
      [arXiv:2001.04201 [nucl-ex]].

\bibitem{Shen:2015saf}
      C.~Shen, Z.~Qiu, U.~Heinz,
      Phys.\ Rev.\ C {\bf 92}, 014901 (2015)
      [arXiv:1502.04636 [nucl-th]].

\bibitem{Niemi:2012edo}
      H.~Niemi, G.~S.~Denicol, H.~Holopainen, P.~Huovinen,
      Phys.\ Rev.\ C {\bf 87}, 054901 (2012)
      [arXiv:1212.1008 [nucl-th]].

\bibitem{Liu:2019pca}
      Z.~Liu, W.~Zhao, H.~Song,,
      Eur.\ Phys.\ J.\ C {\bf 79}, 870 (2019)
      [arXiv:1903.09833 [nucl-th]].

\bibitem{Xiao:2016foe}
      K.~Xiao, F.~Liu, F.~Wang,
      Phys.\ Rev.\ C {\bf 94}, 024905 (2016)
      [arXiv:1509.06070 [nucl-th]].

%\bibitem{Milosevic:2017nro}
%      J.~Milosevic,
%      [arXiv:1708.09717 [nucl-ex]].

%\bibitem{Sirunyan:2019mha}
%      A.~Sirunyan {\it et al.} [CMS Collaboration],
%      [arXiv:1910.08789 [nucl-ex]].

\bibitem{Lin:2004amt}
      Z.-W.~Lin, C.~Ko, B.~A.~Li, B.~Zhang and S.~Pal,
      Phys.\ Rev.\ C {\bf 72}, 064901 (2005)
      [arXiv:nucl-th/0411110 [nucl-th]].

\bibitem{De:2018hri}
      D.-X.~Wei, X.-G.~Huang and L.~Yan,
      Phys.\ Rev.\ C {\bf 98}, 044908 (2018)
      [arXiv:1807.06299 [nucl-th]].

\bibitem{CMS:2017mom}
      CMS, [CMS Collaboration],
      [CMS-PAS-HIN-16-018].

\bibitem{Acharya:2017sft}
      S.~Acharya {\it et al.} [ALICE Collaboration],
      JHEP {\bf 09}, 032 (2017)
      [arXiv:1707.05690 [nucl-ex]].
%%%%%%%%%%%%%%%%%%%%%%%%%%%%%%%%%%%%%%%%%%%%%%%%%%%%%%%%%%%%%%%%%

\end{thebibliography}

\end{document}